\documentclass[12pt]{article}
\usepackage[dvips]{epsfig}
\usepackage{amssymb}
\def\bea{\begin{eqnarray}}
\def\eea{\end{eqnarray}}
\def\nn{\nonumber }

\newcommand{\be}{\begin{equation}}
\newcommand{\ee}{\end{equation}}
\newcommand{\p}{\partial}

\def\no{\noindent}

\def \tf1{\tilde {f}_1}
\def \tf2{\tilde {f}_2}

\def \eps {\epsilon_{\mu\nu\alpha}}
\newcommand{\beq}{\begin{equation}}
\newcommand{\eeq}{\end{equation}}

\begin{document}

\title{\textbf{Static potential in
scalar QED$_3$ with non-minimal coupling}}
\author{ D. Dalmazi and Elias L. Mendon\c ca\\
\it{UNESP - Campus de Guaratinguet\'a - DFQ} \\
\it{Av. Dr. Ariberto Pereira da Cunha, 333} \\
\it{CEP 12516-410 - Guaratinguet\'a - SP - Brazil.} \\
\sf{E-mail:  dalmazi@feg.unesp.br }}
\date{\today}
\maketitle

\begin{abstract}

Here we compute the static potential in scalar $QED_3$ at leading
order in $1/N_f$. We show that the addition of a non-minimal
coupling of Pauli-type ($\eps j^{\mu}\partial^{\nu}A^{\alpha}$),
although it breaks parity, it does not change the analytic
structure of the photon propagator and consequently the static
potential remains logarithmic (confining) at large distances. The
non-minimal coupling modifies the potential, however, at small
charge separations giving rise to a repulsive force of short range
between opposite sign charges, which is relevant for the existence
of bound states. This effect is in agreement with a previous
calculation based on M$\ddot{o}$ller scattering, but differently
from such calculation we show here that the repulsion appears
independently of the presence of a tree level Chern-Simons term
which rather affects the large distance behavior of the potential
turning it into constant.

\it{PACS-No.:} 11.15.Bt , 11.15.-q
\end{abstract}



\newpage

\section{Introduction}


An important problem in high energy physics is the lack of a
rigorous proof of  color confinement in 4D QCD. Different
techniques have been used to tackle this problem. We can mention
lattice simulations \cite{lattice}, supersymmetry \cite{sw} and
lower dimensional models \cite{rrs,gkms,amz}.

In order to investigate the contribution of the matter fields to
this problem we integrate over such fields in the path integral
and derive an effective action for the vector bosons, solving the
equations of motion of this quantum action we can compute the
potential between two static charges separated by a distance $L$.
A monotonically increasing potential as $L\to\infty$ signalizes
confinement. This route has been followed in
\cite{ab,ghosh,addh,prd} in the case of $QED_3$. In that model, if
we work with two-component fermions, the fermion mass term breaks
parity and a Chern-Simons term is dynamically generated leading to
an important change in the analytic structure of the photon
propagator which turns the classically confining logarithmic
potential into a constant at large distances.

In the case of scalar $QED_3$ we have a different scenario since
its mass term, like the rest of the Lagrangian, is parity
symmetric and no parity breaking term is dynamically generated, so
the classical logarithmic potential survives at quantum level.
Therefore, it is expected that the inclusion of parity breaking
terms in the Lagrangian would strongly modify the static
potential. A natural possibility to be considered is a non-minimal
coupling of Pauli-type which breaks parity but preserves gauge
invariance. This term is rather simple in $D=3$ where the dual
field strength $(F_{\mu}=\eps A^{\mu}\p^{\nu}A^{\alpha})$ is a
pseudo-vector and the addition of the non-minimal coupling amounts
to the replacement $ e A_{\mu} \to e A_{\mu} + \gamma F_{\mu}$
where $\gamma $ is the non-minimal coupling constant which has
negative mass dimension. This term has been considered before in
the literature of $QED_3$ and scalar $QED_3$, see e.g. [9-16].

Another motivation for the inclusion of the non-minimal coupling
comes from \cite{ck,i,na} where there are indications, see however
\cite{h}, that the coupling of a gauge field to fermions via a
Pauli term could give rise to anyons with no need of a
Chern-Simons term. Since the change of statistics is a long range
phenomena and the Chern-Simons term indeed changes the static
potential at large distances, we would like to include the
Pauli-type interaction in order to check, at least in some
approximation, if it could really produce large distance effects.

A further point concerns previous calculations in the literature.
It has been claimed in \cite{gw,ghosh2} that the effect of the
non-minimal coupling on the static potential only appears if a
Chern-Simons term is present. This is apparently not the case of
$QED_3$ with four-component fermions where no Chern-Simons term is
generated but still there is some influence of the Pauli-type term
on the static potential at low distances \cite{prd}. It is
important to remark, however that here and in \cite{prd} one works
at leading order in $1/N_f$ which requires the calculation of the
one loop vacuum polarization diagram, while the calculations of
\cite{gw,ghosh2} are based on the one photon exchange diagram at
tree level (Moller scattering) in the non-relativistic limit. In
order to control the effect of the Chern-Simons term and compare
our results to \cite{gw,ghosh2} we introduce here, besides the
Pauli-type term, a Chern-Simons term at tree level with an
arbitrary coefficient.

We have already mentioned that the Pauli term demands a coupling
constant with negative mass dimension (non-renormalizable) so we
found suitable to use $1/N_f$ expansion since there are some
arguments \cite{gms} in favor of the $1/N_f$ renormalizability of
such interaction. In the next section we start by integrating over
the $N_f$ scalar fields at leading order in $1/N_f$. Then, we
analyze the analyticity properties of the corresponding photon
propagator. In section III we minimize the effective action and
compute the static potential $V(L)$ numerically for a finite
scalar mass and analytically in the limit $m\to\infty $. We draw
some conclusions in section IV.

 \section{The photon propagator at $N_f\to\infty$}

Our starting point is to integrate over the $N_f$ scalar fields
$\phi_r$ , $\, r=1,2,\cdots ,N_f $ in the partition function
below:

\begin{eqnarray}
Z &=&\int \mathcal{D}A_{\mu }\, e^{i\,\int \,d^{3}x\, {\cal
L}\left(A^{\mu},j_{\rm ext} ^{\nu }\right)}
\prod_{r=1}^{N_f}\mathcal{D}%
\phi^*_r \,\mathcal{D}\phi_r\, e^{ -\frac i2 \,\int \,d^{3}x
\phi_r^* \left\lbrack D^{\mu}D_{\mu} + m^2 \right\rbrack \phi_r }
\nonumber\\
&=& C\, \int \mathcal{D}A_{\mu }\, e^{i\,\int \,d^{3}x\, {\cal
L}\left(A^{\mu},j_{\rm ext} ^{\nu }\right) - N_f {\rm Tr} \ln
\left\lbrack  D^{\mu}D_{\mu} + m^2 \right\rbrack }\label{z1}
\end{eqnarray}

\no where $C$ is a numerical constant and

\be  {\cal
L}\left(A^{\mu},j_{\rm ext} ^{\nu }\right) = -\frac{1%
}{4}\,F_{\mu \nu }^{2}\, - \frac{\theta}2 \eps
A^{\mu}\partial^{\nu}A^{\alpha} \, + \frac{\zeta }{2}\left(
\partial _{\mu }A^{\mu }\right) ^{2} \, - A_{\nu}j^{\nu}_{\rm ext}
\label{lext} \ee

\no The external current corresponding to a static charge $Q$ at the point
$(x_1,x_2)=(L/2,0)$ is given by  $j_{\rm ext}^{\nu} = Q \delta(x_2)\delta({x_{1}} -
{\frac{L }{2}}) \delta^{\nu 0}$. Later on, the interaction energy of a couple of
charges $-Q$ and $Q$ separated by a distance $L$ will be calculated via $V(L) = -Q A_0
\left(x_1=-L/2,x_2=0\right)$ where $A_0(x_{\nu})$ will be obtained minimizing the
effective action coming from (\ref{z1}). The covariant derivative: $D_{\mu}\phi \, =\,
\partial_{\mu}\phi - i e \,\phi A_{\mu}/\sqrt{N_f} - i\gamma \, \phi F_{\mu} /\sqrt{N_f}$  includes
the non-minimal coupling constant $\gamma$ which has negative mass
dimension $\left\lbrack\gamma\right\rbrack=-1/2$ while
$\left\lbrack e \right\rbrack= 1/2$ and
$\left\lbrack\theta\right\rbrack= 1$. The dual of the strength
tensor is defined here as $F_{\mu}=\eps
\partial^{\nu}A^{\alpha}  $.

The next step is to evaluate the trace of the logarithm
perturbatively in $1/N_f$. We have two types of interaction
vertices coming from ${\cal L}_{int}^{(1)}= i
\left(\phi^*\p_{\mu}\phi - \phi\p_{\mu}\phi^*\right)\left(e
A^{\mu} + \gamma F^{\mu}\right)/\sqrt{N_f}$ and ${\cal
L}_{int}^{(2)}= i \phi^*\phi \left(e A^{\mu} + \gamma
F^{\mu}\right)^2/N_f$. Thus, the leading contribution in $1/N_f$
would come from just one vertex of the first type, however, since
it involves derivatives of the scalar fields the Feynman rules for
scalar $QED$ include a factor $p_{\mu}^{\rm in} + p_{\mu}^{\rm
out}$ where those are the incoming and out-coming momenta of the
scalar fields. Therefore the diagram (tadpole) will be
proportional to the integral $\int d^3p \, (p_{\mu}^{\rm in} +
p_{\mu}^{\rm out})/(p^2-m^2)=\int d^3p \, 2 p_{\mu} /(p^2-m^2)$
which vanishes in the dimensional regularization adopted here. The
next leading contribution includes either two vertices of the
first type or one vertex of the second type. Both contributions
will be independent of $N_f$ due to the overall factor $N_f$ in
front of the logarithm in (\ref{z1}) and will survive the limit
$N_f \to \infty$. The next contribution would come from three
vertices of the first type and would be of order $1/\sqrt{N_f}$ so
it vanishes if $N_f\to\infty$. Such contributions and higher ones
will be neglected henceforth. In conclusion we have, up to an
overall constant, $Z =\int \mathcal{D}A_{\mu }\, e^{i \, S_{\rm
eff}}$ where:

\be S_{\rm eff} \, = \, \int d^3x {\cal L}\left(A^{\mu},j_{\rm
ext} ^{\nu }\right) - \frac i2 \int d^3k \left(e
\tilde{A}^{\mu}(k) + \gamma \tilde{F}^{\mu} (k) \right) T_{\mu\nu}
\left(e \tilde{A}^{\nu}(-k) + \gamma \tilde{F}^{\nu} (-k) \right)
\label{leff} \ee

\no The quantities $\tilde{A}_{\mu},\tilde{F}_{\nu}$ are Fourier
transforms and

\be T_{\alpha\beta}\, = \, -2 g_{\alpha\beta} I^{(1)} \, + \,
I_{\alpha\beta}^{(2)} \label{tab} \ee

\no Using dimensional regularization we have obtained for the
Feynman integrals:

\be I^{(1)} = \int\frac{d^3 p}{(2\pi)^3}\frac 1{p^2-m^2} = i\frac
{m}{4\pi} \label{i1}\ee

\be I^{(2)}_{\alpha\beta} \!=\! \int \frac{d^3 p}{(2\pi)^3}\frac
{(2p + k)_{\alpha}(2p + k)_{\beta}} {(p^2-m^2)\left\lbrack (p+
k)^2 - m^2\right\rbrack } = \frac{i m}{8\pi}\left\lbrack 4
g_{\alpha\beta} - 2 z f_2
\theta_{\alpha\beta}\right\rbrack \\
\label{i2}\ee

\no With $z=k^2/4 m^2$ and $\theta_{\alpha\beta} = g_{\alpha\beta}
- k_{\alpha}k_{\beta}/k^2$. In the region $0 \le z \le 1$ we have

 \bea f_2 \, &=& \, \frac 1z \left\lbrack 1 +
\frac{1-z}2 f_1
\right\rbrack \, = \, \frac 23 + \frac 2{15} z + \frac{2}{35} z ^2 + \cdots \label{f2}\\
f_1 \, &=& \, - \frac{1}{\sqrt{z}}\ln \frac{1 + \sqrt{z}}{1 -
\sqrt{z}} \label{f1} \eea

Above the pair creation threshold $( z > 1 )$ the integral
$I^{(2)}_{\alpha\beta}$ develops a real part which will be
neglected henceforth. For future use we have given the large mass
expansion for $f_2$. The static potential requires the expression
for the effective action for $z < 0$ which can be obtained by
analytically continuing (\ref{f2}) and (\ref{f1}). Namely, with
$\tilde{z} = -z
> 0 $ we have :

\bea \tilde{f}_2 \, &=& \, -\frac 1{\tilde{z}} \left\lbrack 1 +
\frac{1+\tilde{z}}2 \tilde{f}_1 \right\rbrack \, = \, \frac 23 -
\frac 2{15} \tilde{z} + \frac{2}{35} \tilde{z}^2 +
 \cdots \label{f2t}\\
\tilde{f}_1 \, &=& \, - \frac{2}{\sqrt{\tilde{z}}}
\arctan\sqrt{\tilde{z}} \label{f1t} \eea

\no  Our result for $T_{\alpha\beta}$ is in agreement with
\cite{AGS} and it is transverse
$k^{\alpha}T_{\alpha\beta}=0=T_{\alpha\beta}k^{\beta}$ in
accordance with gauge invariance. Now we can write down the
effective action for scalar $QED_3$ including vacuum polarization
efffects :

\bea S_{\rm eff} \, &=& \, \int d^3x\, \left\lbrace -\frac 14
F^{\mu\nu}\left\lbrack 1 - \frac{\gamma^2 \Box f_2}{16 \pi m} +
\frac{e^2 f_2}{16 m \pi} \right\rbrack F_{\mu\nu} + \frac{\zeta
}{2}\left(
\partial _{\mu }A^{\mu }\right)^{2} \right. \nonumber\\
&-& \left. \frac{\theta}2 \eps A^{\mu}\partial^{\nu}A^{\alpha} -
\frac{e\gamma}{16 m \pi} \eps A^{\mu}\partial^{\nu}\Box f_2
A^{\alpha} - A_{\nu}j^{\nu}_{\rm ext} \right\rbrace\label{seff}
\eea

\no where $f_2=f_2(-\Box/4 m^2)$ is given in (\ref{f2}) and
(\ref{f2t}). Notice that, besides the tree level Chern-Simons
term, another  parity breaking term appears in (\ref{seff})  due
to the magnetic moment interaction. Although the action
(\ref{seff}) is non-local it can be made local in the large mass
limit $m\to\infty$ as in \cite{fs}. Introducing the dimensionless
constants

\be c_1 = \frac{e^2}{16\pi m} \, ; \, c_2 = \frac{e\gamma}{8\pi}
\, ; \, c_3 = \frac{\theta}{2m} \label{constants} \ee

\no Taking  $m\to\infty $ while keeping the dimensionless
constants finite, the only effect of the vacuum polarization is a
finite renormalization of the Maxwell term, i.e.,

\be S_{\rm eff}\left(m\to\infty\right) = \, \int d^3x\,
\left\lbrack -\frac {1 + 2 c_1/3}4 F_{\mu\nu}^2 - \frac{\theta}2
\eps A^{\mu}\partial^{\nu}A^{\alpha} + \frac{\zeta }{2}\left(
\partial _{\mu }A^{\mu }\right) ^{2} - A_{\nu}j^{\nu}_{\rm ext}
\right\rbrack \label{seffm}\ee

\no On the other hand, for finite mass we can write down:

\bea S_{\rm eff} &=& \int d^3 x d^3y \left\lbrack
A^{\mu}(x)\frac{D^{-1}_{\mu\nu}(x,y)}2 A^{\nu}(y)  -
A_{\nu}j^{\nu}_{\rm ext}\delta^{(3)}(x-y)\right\rbrack
\label{seffxy} \\
&=& \int \frac{d^3
k}{(2\pi)^3}\tilde{A}^{\mu}(k)\frac{\tilde{D}_{\mu\nu}(k)^{-1}}2\tilde{A}^{\nu}(-k)
- \int d^3 x A_{\nu}j^{\nu}_{\rm ext} \label{dk} \eea

\no Where the photon propagator in momentum space is given by

\be \tilde{D}_{\mu\nu} = a \, \left(g_{\mu\nu} -
\theta_{\mu\nu}\right) + b \, \theta_{\mu\nu} + c \, \eps
k^{\alpha} \label{propa}\ee

\no with

\be a  = \frac 1{\zeta k^2} \label{a} \ee

\be  b = - \frac{c_1\left(D_+ + D_-\right)}{8 m^2 \sqrt{z} D_+
D_-} \label{a+b} \ee

\be  c = - i \frac{c_1\left(D_+ - D_-\right)}{16 m^3 z D_+ D_-}
\label{c} \ee

\be D_{\pm} \, = \, \sqrt{z}\left\lbrack c_1 + \left(c_1 \pm
\sqrt{z} c_2\right)^2 f_2 \right\rbrack \pm c_1 c_3 \equiv
g_{\pm}(z) \pm c_1 c_3 \label{dpm} \ee

\no Now we are able to analyze the analyticity properties of the
photon propagator. First of all, we notice that the massless pole
$z=0$  in the denominator of (\ref{c}), which is typical of a
Maxwell-Chern-Simons theory, is a gauge artefact. It disappears
from gauge invariant correlators involving the field strength
$F_{\mu}$. It can be shown \cite{Baeta1} to have a vanishing
residue (non-propagating mode). Since the factor $\sqrt{z}$ in the
denominator of (\ref{a+b}) is cancelled ed out by the numerator,
the only possibilities for poles in the propagator stem either
from $D_+=0$ or $D_- = 0$. Due to $c_1 > 0$ and $2/3 \le f_2 < 1$
we have $ g_{\pm}(z)> 0 $ and consequently we can only have
$D_+=0$ or $D_- = 0$ for $c_3 < 0$ or $c_3
> 0$ respectively. We never have two poles at the same time.
In the absence of the Chern-Simons term, i.e., $c_3=0$, the
product $D_+ D_-$ will be proportional to $z$ and we are left with
one massless pole $z=0$. Since $\lim_{z\to 0} z (a + b) = -
1/\left(1 + 2 c_1/3\right) < 0$ the residue at this pole will be
positive and this represents a physical massless photon which will
be responsible for a long range logarithmic static potential. On
the other hand, if $c_3 \ne 0 $, since the denominator $D_+ D_-$
is symmetric under $c_2 \to - c_2 \, ; \, c_3 \to - c_3$, it is
enough to consider only $D_+=0$ assuming $c_3 < 0$ the other case
$D_-=0$ with $c_3
> 0$ follows from the symmetry. Numerically, we have checked that whatever sign we
choose for $c_2$ the function $g_+(z)$ is monotonically increasing
and satisfies $g_+(z) > 0 $ consequently its maximum is $g_+(1)$.
Therefore, see (\ref{dpm}), if $c_3 < - g_+ (1)/c_1 = \left\lbrack
(c_1 + c_2)^2 + c_1\right\rbrack /c_1 $ then we have no poles and
so no particle in the spectrum. On the opposite, if $ - g_+
(1)/c_1 < c_3 < 0 $ we are always able to find numerically one
massive pole for some  $0 < z < 1$ such that $D_+=0$ which is a
typical effect of a Chern-Simons term \cite{djt}. As we move
toward the left limit value $c_3 \to - g_+ (1)/c_1$, the photon
mass increases to the point where it reaches the pair creation
threshold $k^2=4m^2$ at $c_3=- g_+ (1)/c_1$. Due to the symmetry
$c_2 \to - c_2 \, ; \, c_3 \to - c_3$ we conclude that whenever
the tree level Chern-Simons term is present and its coefficient is
not too much negative or too much positive ($\vert c_3 \vert <
g_+(1)/c_1 $) we have one massive physical (positive residue)
photon and if the Chern-Simons  term is absent we have one
physical massless photon.

It is remarkable to find a no poles region in the photon
propagator. One might think that this is due to some convergence
problem of the $1/N_f$ expansion which has been introduced because
of the non-renormalizable non-minimal coupling $c_2$. However,
even if $c_2=0$ the Chern-Simons coefficient must obey an upper
bound $\vert c_3 \vert < g_+(1)/c_1 = 1 + c_1 $ in order to have a
physical pole in the photon propagator at one loop level. By
analytically continuing, see (\ref{f2t}), the expression for the
propagator to the region $z = k^2/4m^2 < 0$ we have checked that
tachyons can only appear for a special fine tuning of the coupling
constants for which we did not find any special interpretation,
namely, the tachyonic pole must be a solution of
$\tilde{z}\tilde{f}_2 = c_3/(2 c_2) $ and this solution must be
such that $c_1^2 c_3 = - \tilde{z} c_2 (2 c_1 - c_2 c_3)$,
although explicit numerical solutions are possible we have found
those fine tuned cases rather artificial. In particular, they have
apparently no relationship with the no-pole region ($\vert c_3
\vert > g_+(1)/c_1 $) and will be disregarded in this work. In the
next section we use the photon propagator as an input to calculate
the static potencial $V(L)$.

\section{The static potential $V(L)$ }

Minimizing the effective action (\ref{seffxy}) we obtain:

\bea A_{\beta}(y) \, &=& \, \int d^3 x D_{\beta\alpha}(y,x) j_{\rm
ext}^{\alpha} (x) \\
& = &\int \frac{d^3k}{(2\pi)^3}\tilde{D}_{\beta\alpha}(k)\int d^3x
\, e^{i k\cdot (y-x)}j_{\rm ext}^{\alpha} (x) \label{abeta} \eea

\no  Since the external current is time independent, in
(\ref{abeta}) there will be a factor $\int dx_0 e^{-i k_0 x_0} =
2\pi \delta (k_0)$ which allows an exact integration over $k_0$,
implying $k^{\mu}k_{\mu} = - k_1^2 - k_2^2 = - \vec{k}^2 < 0$,
consequently $z < 0$ and we need the analytic continued functions
$\tilde{f}_2 $ instead of $f_2$. The integrals over $x_1$ and
$x_2$ can also be readily done using the delta functions in the
external current. The angle part of the integral $dk_1 dk_2 = k dk
d\theta $ gives rise to the Bessel function $J_0 (k L)$. Thus, we
are left with the radial integral over $k=\sqrt{k_1^2 + k_2^2}$.
Placing the negative charge $-Q$ at $(y_1,y_2)=(-L/2,0)$ we have

\bea V(L) &\, =\, & \, -Q \, A_0(y_1=-L/2,y_2=0) \nn\\
&\, = \, & -\frac{Q^2}{2\pi} \int_{0}^{\infty} dk \, k \,
\tilde{b}\, J_{0}(k L) \label{vl} \eea

\no  The tilde in the expression $\tilde{b}$ stands for the
analytic continuation of (\ref{a+b}) to $z<0$.

Now we discuss some special cases starting with the pure scalar
$QED_3$ where $c_3=0=c_2$. In this case $k \tilde{b}\, J_0(k L) =
J_0(k L)/\left\lbrack k (1 + c_1 \tilde{f}_2)\right\rbrack $ since
$\tilde{f}_2(k=0)=2/3$ and $J_0(0)=1$ we have an infrared
divergence at $k=0$ and the integral (\ref{vl}) is divergent as it
stands. We make a subtraction in order to get rid of this infrared
divergence and define:

\be V(L) - V(L_0) =  - \frac{Q^2}{2\pi} \lim_{x\to 0}
\int_{x}^{\infty} dk \, k \, \tilde{b} \left\lbrack J_{0}(k L) -
J_0 (k L_0) \right\rbrack \label{vll0} \ee

\no In general the integral (\ref{vll0}) must be calculated
numerically, one exception is the large mass limit $m\to\infty$.
In this case $k \tilde{b} \to 1/\left\lbrack k (1 + 2
c_1/3)\right\rbrack $ and the integral can be calculated exactly
\cite{as}:

\be \left\lbrack V(L) - V(L_0)\right\rbrack_{m\to\infty} =
\frac{Q^2_{\rm R}}{2\pi} \ln\left(\frac L{L_0}\right)
\label{vll0net}\ee

\no where

\be Q_{\rm R} = \frac Q{\left\lbrack 1 + \frac{e^2}{24\pi
m}\right\rbrack^{1/2}} \quad . \label{qnet} \ee

\no The classical potential is given by (\ref{vll0net}) with
$Q_{\rm R}$ replaced by the bare charge $Q$. Therefore, the sole
effect of the vacuum polarization is a finite renormalization of
the charge. The situation is similar to $QED_3$ with
four-component fermions where no Chern-Simons term is dynamically
generated, the only difference is the renormalized charge which is
$Q/\left\lbrack 1 + e^2/(6\pi m) \right\rbrack $ instead of
(\ref{qnet}). Thus, the renormalization factor is larger for
fermions than for scalars. For finite mass the potential must be
calculated numerically. We plot\footnote{In all figures in this
work the symbol $V$ stands actually for the difference $
V(L)-V(L_0)$} the results in figure 1 for the masses $m=1$ and
$m=0.01$ and compare with the classical result and the result of
\cite{prd} for four-component fermions. We notice that the finite
renormalization due to the vacuum polarization is always stronger
for fermions than for scalars and its effect increases with the
mass of the matter fields. For both scalar $QED_3$ and $QED_3$ we
see in figure 1 that the numerically calculated static potential
at $m=1$ is already very close to the analytic result (solid
lines) obtained in the limit $m\to\infty$.

\begin{figure}
\begin{center}
\epsfig{figure=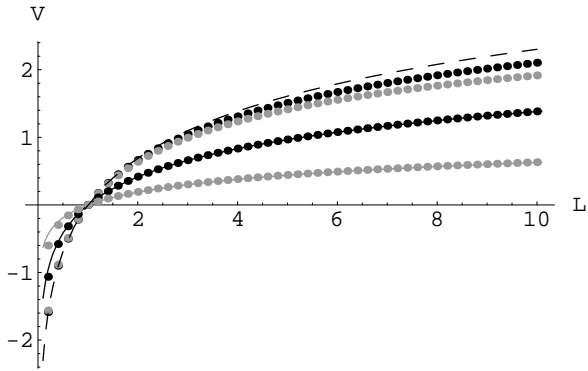,width=80mm} \caption{The static
potential for pure scalar $QED_3$ (dark dots) and pure $QED_3$
with four-component fermions (light dots). The dashed line
corresponds to the classical potential. We have fixed
$(c_1,c_2,c_3)=(1,0,0)$ and $m=0.01$ for the two dotted curves
closer to the classical potential while we have $m=1$ for the
farther curves which overlap with the $m\to\infty$ analytic result
(solid curves)} \label{figure1}
\end{center}
\end{figure}

Next, we check the effect of the non-minimal coupling $c_2\ne 0$
in the absence of the Chern-Simons term $(c_3=0)$. In figure 2 we
see that for $L\to\infty$ the effect of the non-minimal coupling
in the vacuum polarization disappears and the potential becomes
logarithmic which can be explained technically by the fact that
the Bessel function $J_0(k L)$ oscillates with decreasing
amplitude as $L\to\infty$ and so the integral will be dominated by
the pole at the origin $k=0$ which makes the higher derivative,
see (\ref{seff}), contribution of the non-minimal coupling
negligible. However, in a finite range close to $L=0$ the
non-minimal coupling gives rise to a surprising repulsive force in
a much similar way to what happens in the case of four-component
fermions in \cite{prd}. Such repulsive force may play an important
role in the existence of bound states. Differently from the
calculation based on the M$\ddot{o}$ller scattering \cite{gw} we
see here effects of the non-minimal coupling even in the absence
of the Chern-Simons term.

\begin{figure}
\begin{center}
\epsfig{figure=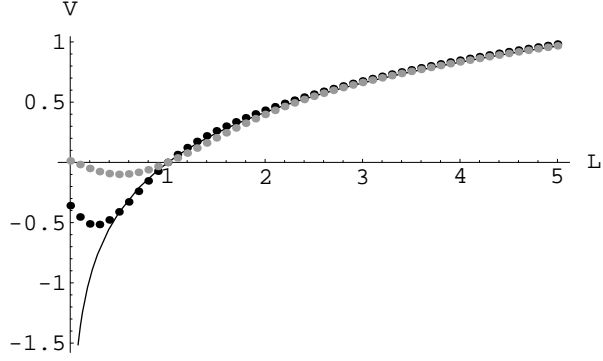,width=80mm} \caption{The static
potential for scalar $QED_3$ without the tree level Chern-Simons
term ($c_3=0$). The solid line corresponds to pure scalar $QED_3$
($c_2=0$) while the dark (light) dots to $c_2=2 (c_2=4)$. We have
assumed $m=3$ and $c_1=1$.} \label{figure2}
\end{center}
\end{figure}

\begin{figure}
\begin{center}
\epsfig{figure=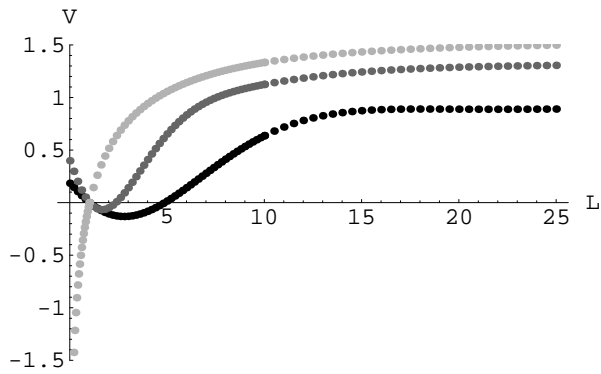,width=80mm} \caption{The static
potential for scalar $QED_3$ with the tree level Chern-Simons term
($c_3=1$). From the lightest to the darkest curve we have $c_2 = 0
; 0.4; 0.8$. We have assumed $m=1=c_1$.} \label{figure3}
\end{center}
\end{figure}

Now we turn on the Chern-Simons term ($c_3\ne 0$). As we see in
figure 3, the potential $V(L) - V(L_0)$ tends to the  constant
$-V(L_0)$ as $L\to\infty$ like the case of  $QED_3$ with
two-component fermions, see \cite{ab}, where a Chern-Simons term
is dynamically generated. Once again, a repulsive force appears
for small separations as we switch on the non-minimal coupling. As
$L\to\infty$ the only effect of the non-minimal coupling is to
change the constant $-V(L_0)$. Although, the plot in figure 3 has
been made for $m=1$ and $c_1=1$ we have checked that the same form
of the potential persists for other values of those constants. In
summary, the effect of the non-minimal coupling is qualitatively
the same in the presence of a Chern-Simons term.

\section{Discussions and Conclusion}

In the case of $QED_3$ with two-component fermions it is well
known that a Chern-Simons term is dynamically generated which
makes the photon massive and turns the classical confining
logarithmic  potential into constant at large distances. In scalar
$QED_3$ the mass term for the scalars is parity invariant and no
Chern-Simons term is dynamically generated and so the classical
logarithmic potential survives vacuum polarization effects. Here
we have explicitly confirmed that fact and analyzed the effect of
adding a parity breaking non-minimal coupling term of Pauli-type
as well as a tree level Chern-Simons term. It turns out that the
non-minimal coupling by itself neither affects the analytic
properties of the photon propagator nor changes the large distance
behavior of the static potential which is unexpected from the
point of view of the interpretation that this term may originate
anyons with no need of a Chern-Simons term, see \cite{ck,na,i} but
see also \cite{h}. On the other hand, at small charge separations
the non-minimal coupling gives rise to a repulsive force between
opposite sign charges which has been observed before in \cite{gw}
by computing the one-photon exchange diagram (M$\ddot{o}$ller
scattering) and taking the non-relativistic limit.
Notwithstanding, the effect found in \cite{gw} only appears in the
presence of the tree level Chern-Simons term  and its attractive
or repulsive nature depends on the sign of $(1-\gamma\theta/e)$
contrary to what we have found here where the non-minimal coupling
influence is present, see figure 2, even if $\theta =0 $ and its
effect is always repulsive independently of the sign of $\gamma$
or $\theta$.

Concerning the tree level Chern-Simons term, as expected, it gives
mass to the photon and shifts the zero momentum pole in the
integral involved in the static potential (\ref{vll0}). The
absence of a singularity in the integration path allows us to take
the limit $L\to\infty$ before performing the integral and so it
will vanish as a consequence of $J_0(x\to\infty)\to 0$. This
effect of the Chern-Simons term was certainly not surprising.
However, it is remarkable that we found an upper bound for the
absolute value of the Chern-Simons coefficient in order to have a
physical pole in the photon propagator at one loop level. As we
increase such absolute value the photon mass increases and
penetrates the real pair creation region $k^2 \ge 4 m^2$ for
finite values of the coupling constants of the theory. We can
mention that this situation is not peculiar to scalar fields since
we have noticed in \cite{addh} that it happens also in $QED_3$
with two-component fermions. In that case, if $c_1 = e^2/(16\pi m)
\ge 1$ there will be no poles in the photon propagator at one loop
level. However, one could argue that $c_1$ is a dimensionless
constant which controls the perturbative expansion (for $N_f=1$)
such that the upper bound could be understood as a limit for
perturbation theory. This argument does not work for scalar
$QED_3$ even if we drop the non-minimal coupling ($c_2=0$) since
the upper bound increases with $c_1$ which makes the latter case
more intriguing.

 At last, we
notice that the static potential in pure scalar $QED_3$ without
tree level Chern-Simons term has been studied in \cite{dz} where
the authors conclude that the potential is of screening type and
even fractional charges can be fully screened. However, the
authors of \cite{dz} have neglected terms of order $e^2/m$ which
have been considered here. Besides, they have gone above the pair
creation threshold and made use of variational methods altogether
with a peculiar Ansatz for the two particle wave function, so we
can hardly compare their findings with our results obtained below
the pair creation threshold.

\section{Acknowledgments}

This work was partially supported by CNPq, FAPESP and CAPES-PROAP,
Brazilian research agencies. We thank Luiz Claudio de Albuquerque,
Marcelo Gomes and Marcelo Hott for discussions.

\end{document}